\documentclass[12pt,a4paper]{article}
\usepackage{jheppub}
\usepackage[T1]{fontenc}
\usepackage{slashed}
\usepackage{caption}
\usepackage{subcaption} 
\usepackage{float}

\DeclareMathOperator*{\res}{Res}

\usepackage{empheq}%

\usepackage{comment}

\usepackage{xcolor}
\usepackage{color,soul}

\usepackage{bbold}

\usepackage{graphicx}

\usepackage{feynmp}

\def\Beq{\begin{equation}}
\def\Eeq{\end{equation}}

\def\Bal{\begin{aligned}}
\def\Eal{\end{aligned}}

\def\[{\left[}
\def\]{\right]}
\def\({\left(}
\def\){\right)}

\newcommand{\be}{\beta}

\def \be {\begin{equation}}
\def \ee {\end{equation}}
\def \bea {\begin{eqnarray}}
\def \eea {\end{eqnarray}}
\def \beal#1 {\begin{align}#1\end{align}}

\everymath{\displaystyle}

\preprint{}

\title{A duality in string theory on $\text{AdS}_3$}

 \author[a]{Indranil
  Halder\note{ihalder@g.harvard.edu},}  \author[a]{Daniel Louis Jafferis\note{jafferis@g.harvard.edu},} \author[a] {David K. Kolchmeyer\note{dkolchmeyer@g.harvard.edu} }

\affiliation[a]{Center for the Fundamental Laws of Nature, Harvard University, Cambridge, MA, USA}

\abstract{We consider bosonic string theory on $\text{AdS}_3$ supported by Kalb-Ramond flux. It is well known that the $\alpha'$ exact worldsheet theory is described by the $\text{SL(2,R)}$ WZW model. In this note, we perform checks of an $\alpha'$ exact dual description proposed in [arXiv:2104.07233] involving a winding condensate on a  free theory background. We give the explicit map of vertex operators for normalizable states on both sides of the duality and demonstrate the equivalence of their two and three-point functions by direct computation. The duality is of strong-weak nature in $\alpha'$.  }

\begin{document}

\maketitle

\section{Introduction}

In this paper, we confirm a strong-weak duality of two-dimensional conformal field theories, between the AdS$_3$ sigma model with Kalb-Ramond flux and a free theory deformed by a winding condensate. This duality, proposed in \cite{Jafferis:2021ywg}, can be understood as an AdS$_3$ uplift of the FZZ duality that relates the cigar theory with sine-Liouville and is related to the duality discussed in \cite{Berkooz:2007fe}. Viewing AdS$_3$ as a solid cylinder, in the dual description the boundary circle is non-contractible, and winding around that direction is violated explicitly by the condensate. 

Strings in AdS$_3$ have a worldsheet conformal field theory given by a non-linear sigma model including Kalb-Ramond $H_3$ flux. The worldsheet CFT has quantum corrections in a $1/k$ expansion, where $k$ sets the curvature radius of AdS in string units $l_{AdS}=\sqrt{\alpha' k}$. Consider the target space for large $k$ in AdS-Rindler coordinates, corresponding in the Euclidean theory to a solid cylinder.
\begin{equation}
	\begin{aligned}
		ds^2_{AdS-Rindler}=l_{AdS}^2(-\sinh^2(\hat{r})d\hat{t}^2+\cosh^2(\hat{r})d\hat{\xi}^2+d\hat{r}^2)
	\end{aligned}
\end{equation}
An essential feature of this theory is that winding around the Euclidean time circle $\hat{\theta}=i \hat{t}$ is not conserved, since it is contractible and strings can unwrap off the tip. This feature is the Euclidean analog of the existence of a coordinate horizon in the Lorentz signature. The idea of the duality is to replace this geometric contractibility with an explicit winding condensate, proceeding as follows. 

The non-linear sigma model is strongly interacting at the AdS boundary since the metric warp factor grows there. Thus it is convenient to pass to the first order description, introducing worldsheet 1-forms $\hat{\chi},\hat{\bar{\chi}} $ living in the target space cotangent bundle. The worldsheet CFT is now described by the following classical action
\begin{equation}\label{actint}
	\begin{aligned}
		& S_{cl}=\frac{k}{2 \pi}\int 2 d^2 \sigma \(\partial \hat{r} \bar{\partial} \hat{r}+\partial \hat{\xi} \bar{\partial} \hat{\xi} +  \hat{\bar{\chi}}\partial (\hat{\xi} -i \hat{\theta})+\hat{\chi} \bar{\partial} (\hat{\xi} +i \hat{\theta})-\frac{1}{\sinh^2  \hat{r}}\hat{\chi}\hat{\bar{\chi}} \)\\
	\end{aligned}
\end{equation}
 The dual proposed in \cite{Jafferis:2021ywg} is obtained by deforming the asymptotic theory obtained by dropping the $\hat{\chi}\hat{\bar{\chi}}$ term in (\ref{actint}) by an on-shell winding operator with zero momentum around the time circle. At the quantum level, this procedure in addition generates a linear dilaton in the radial direction (see (\ref{winding2}) for a detailed presentation). This matches all of the symmetries of global Euclidean AdS$_3$, and the duality reduces to the FZZ duality upon gauging translations along the spatial direction $\hat{\xi}$.

The CFT discussed above can be used for the critical bosonic string, taken together with an internal CFT to achieve $c=26$. The associated string duality is interesting for a variety of reasons that we now review. 

The first is that there is a striking similarity between the condensate worldsheet CFT discussed above, and the AdS/CFT target space dual proposed in  \cite{Eberhardt:2021vsx} (see also \cite{Martinec:2021vpk, Balthazar:2021xeh}) in continuation of the recent developments in \cite{Gaberdiel:2018rqv, Eberhardt:2018ouy, Eberhardt:2019ywk, Dei:2020zui, Gaberdiel:2020ycd, Eberhardt:2020bgq}. More precisely the spacetime CFT proposed in \cite{Eberhardt:2021vsx} is a symmetric product orbifold (of a linear dilaton coupled with the internal CFT) deformed by a marginal operator of twist two. The linear dilaton ($\varphi$) part  of the marginal operator associated with the deformation in the untwisted sector is given by
\begin{equation}
	e^{\sqrt{k-2}\varphi}
\end{equation}
This matches the radial dependence of the winding potential in the condensate worldsheet CFT mentioned above. This presents the possibility of obtaining the spacetime CFT from the worldsheet of the long strings as speculated in \cite{Seiberg:1999xz}.

To explain the second motivation we consider two different analytic continuations to Lorentzian signature. Continuing $\hat{\xi}$ to time leads to a conventional string duality between global Lorentzian AdS$_3$ and a spatial winding condensate which sews together the spatial annulus topology of the free theory. The Lorentzian continuation of $\hat{\theta}$ to time is more subtle, and gives two-sided AdS-Rindler on one hand, and a condensate on top of a disconnected Lorentzian target on the other. In this sense, it is a version of the ER=EPR correspondence between the connected Einstein-Rosen bridge and an entangled superposition of disconnected geometries.

 In the present context, the general Lorentzian formalism developed in \cite{Jafferis:2021ywg} allows us to view the winding condensate around the free theory as a set of entangled states of extended strings emanating from the strong coupling region in the target space for $k>3$ as explained below. This confirms from a complementary perspective the observation that time winding condensates in the related FZZ case leads to the appearance of folded strings in the Lorentzian theory \cite{Giveon:2016dxe,  Itzhaki:2018glf, Giveon:2019gfk, Giveon:2020xxh}. 
To find the Lorentzian continuation, first one sees that the analog of the $i\epsilon$ prescription for Euclidean time winding operators is that the worldsheet is continued to Lorentz signature in Rindler coordinates around the insertion, rather than in radial quantization \cite{Witten:2013pra, deLacroix:2017lif}. This implies that such winding operators produce a state in an angularly quantized Hilbert space of extended strings with asymptotics defined by the operator \cite{Agia:2022srj}. 
Moreover, winding conservation in the free background implies that winding operators must appear in pairs. The result is that the state produced by the worldsheet path integral is the thermofield double state of the extended string Hilbert space. The anomalous momentum conservation in the linear dilaton direction leads to a condensate of such strings, in the sense that their number grows without bound in the perturbative approximation as one approaches the strong coupling infinity.

The goal of this paper is to calculate the correlation functions of all vertex operators of normalizable states (without spectral flow) on the winding condensate side and check their equality with the known $SL(2,C)_k/SU(2)$ correlators. The method we use is developed in \cite{PhysRevLett.66.2051, Dorn:1994xn, Zamolodchikov:1995aa} for Liouville\footnote{See  \cite{Hikida:2008pe, Fukuda:2001jd, Giribet:2021cpa} for a similar discussion in the context of sine-Liouville.}, is to first integrate out the zero mode of the radial direction, imposing its anomalous momentum conservation law due to the linear dilaton. The result is the insertion of a (generally irrational) power of the integrated winding operator into a correlation function of the free asymptotic theory, similar to the insertion of a power of the so-called screening operator in Minimal models. This can be evaluated by analytic continuation from special choices of the parameters where the power is a positive integer\footnote{In this case an even integer since there must be equal numbers of positive and negative winding number operators.}, where the correlator gives the residue at a pole. One can express the residue at these poles in terms of a Liouville theory correlation function (with additional integrations). It is well known that in the sigma model description, the analog of these residues is captured by a Liouville theory with a different set of parameters \cite{Ribault:2005wp}. These two Liouville descriptions turn out to be related to each other under the celebrated $b \rightarrow 1/b$ Liouville duality.

In section \ref{Sigma model description} we review the sigma model description of the worldsheet CFT, in section \ref{Winding condensate description} we present the dual description in terms of the winding condensate, we devote section \ref{Proving the equivalence} to show the equivalence of these two descriptions, we present a brief summary and discuss some open question in section \ref{Summary and open questions}.

\section{Sigma model description}\label{Sigma model description}

The Euclidean worldsheet CFT of strings moving in Euclidean AdS$_3$ (we are not presenting the internal CFT X associated with the compact directions) is somewhat defined by \cite{Giveon:1998ns, Kutasov:1999xu}\footnote{Oftentimes this model is referred to as the $H_3^+$ model.}
\begin{equation}\label{poincare_1st}
	\begin{aligned}
		S=\frac{1}{ \pi}\int  d^2 \sigma \(\frac{1}{2}\partial \phi \bar{\partial} \phi-\frac{1}{4 \alpha_+}\phi R -B \bar{\partial} \Gamma -\bar{B} \partial \bar{\Gamma}-B\bar{B}e^{-\frac{2}{\alpha_+}\phi} \)
	\end{aligned}
\end{equation}
The radius of AdS$_3$ in string units ($\alpha'=1$) is set by $k$ through the formula $$ \alpha_+:=\sqrt{2(k-2)}$$
After integrating out $B$ from the action presented in (\ref{poincare_1st}) we get the usual sigma model description (see (\ref{sigma_model}) for instance).
Relevant OPEs are
\begin{equation}
	\begin{aligned}
		B(z)\Gamma(0)\sim \frac{1}{z}, ~~~ \phi(z, \bar{z})\phi(0,0)\sim -\ln z-\ln \bar{z}
	\end{aligned}
\end{equation}
The conserved currents that we will need are given by\footnote{Expression for the currents are normal ordered, i.e., say $B\partial \Gamma=[B\partial \Gamma]$, etc. Where the normal ordering is defined by $$[AB](w)=\frac{1}{2\pi i }\oint \frac{dz}{z-w} A(z)B(w)\neq [BA](w)$$ }
\begin{equation}
	\begin{aligned}
		& T=-\frac{1}{2} (\partial \phi)^2-\frac{1}{\alpha_+}\partial^2 \phi +B \partial \Gamma\\
		& J^3=\frac{\alpha_+}{2}\partial\phi+\Gamma B
	\end{aligned}
\end{equation}

The delta function normalizable states of the worldsheet CFT are described by vertex operators given by\footnote{The precise numerical factor of proportionality will be fixed later in section \ref{SMcf}.}
\begin{equation}\label{continuous}
	\begin{aligned}
		& V^0_{j,m,\bar{m}} \sim \ \Gamma^{j+m} \bar{\Gamma}^{j+\bar{m}}e^{\frac{2}{\alpha_+}j \phi} , ~~ j=-\frac{1}{2}+is, ~~ s \in \mathbb{R}
	\end{aligned}
\end{equation}
This should be multiplied by a vertex operator associated with the internal CFT, chosen such that overall the vertex operator has worldsheet conformal dimension $(1,1)$.
Here $j,m,\bar{m}$ refers to the charges associated with worldsheet $SL(2,R)$ current algebra. In particular, the OPE of these normalizable operators are known to converge and close among themselves. Therefore the normalizable sector of the worldsheet conformal field theory is entirely described by the two and three-point function of these operators. For this note, we will focus only on these operators.\footnote{The Euclidean worldsheet CFT is formally the SL(2,$\mathbb{C}$)$_k$/SU(2)  WZW model that can be studied exactly in $k$ because the entire effect of the interaction (for a special class of correlation functions, with the rest defined through analytical continuation) is captured in the free theory if we insert a suitable (finite number of) screening operators while computing correlation functions.\footnote{Similar technique known as Coulomb gas formalism has been applied to study unitary minimal models.} }  

To describe the physics of Lorentzian AdS$_3$ we need to consider a more general class of vertex operators obtained by analytical continuation of correlation functions containing vertex operators presented in (\ref{continuous}) to the following  real values of $j$
\begin{equation}\label{discrete}
	-\frac{1}{2}<j<\frac{k-3}{2}
\end{equation}
We would further need to consider spectrally flowed states obtained from (\ref{continuous}) (long strings), as well as  (\ref{discrete}) (short strings) \cite{Maldacena:2000hw, Maldacena:2000kv, Maldacena:2001km}. None of those states apart from the ones in (\ref{continuous})  are normalizable. However, the non-normalizability is restricted to a given point in the spacetime CFT (demanding this gives the lower bound in (\ref{discrete}), and the upper bound is necessary to make sure the contribution of worldsheet instantons is finite).

\section{Winding condensate description}\label{Winding condensate description}

It's easy to see that the fields in (\ref{poincare_1st}) reproduce the Poincare patch of AdS$_3$ by classically integrating out $B,\bar{B}$ at large k to obtain
\begin{equation}\label{sigma_model}
	\begin{aligned}
		& S_{cl}=\frac{k}{ \pi}\int  d^2 \sigma \(\partial \hat{\phi} \bar{\partial} \hat{\phi} +\bar{\partial} \hat{\Gamma} \partial \bar{\hat{\Gamma}}e^{2\hat{\phi}} \)\\
		& \phi:=\alpha_+ \hat{\phi}, ~~ \Gamma:=\frac{\alpha_+}{\sqrt{2}}\hat{\Gamma}, ~~ \bar{\Gamma}:=\frac{\alpha_+}{\sqrt{2}}\hat{\bar{\Gamma}}
	\end{aligned}
\end{equation}
Corresponding formulae at large $k$ in global co-ordinates
\begin{equation}
	\begin{aligned}
		& \hat{\phi}:=-\hat{\xi}+\log \cosh \hat{r},~~ \hat{\Gamma}:=\tanh \hat{r} \ e^{\hat{\xi}+i \hat{\theta}},~~ \hat{\bar{\Gamma}}:=\tanh \hat{r} \ e^{\hat{\xi}-i \hat{\theta}}
	\end{aligned}
\end{equation}
take the following form 
\begin{equation}\label{globalAdS}
	\begin{aligned}
		& S_{cl}=\frac{k}{ \pi}\int  d^2 \sigma \(\partial \hat{r} \bar{\partial} \hat{r}+\partial \hat{\xi} \bar{\partial} \hat{\xi} +\sinh^2  \hat{r} \  \partial (\hat{\xi} -i \hat{\theta})\bar{\partial} (\hat{\xi} +i \hat{\theta})+ i \tanh \hat{r} (\partial \hat{r} \bar{\partial} \hat{\theta}-\partial \hat{\theta} \bar{\partial} \hat{r})\)\\
		& 
	\end{aligned}
\end{equation}
The term $ i \tanh \hat{r} (\partial \hat{r} \bar{\partial} \hat{\theta}-\partial \hat{\theta} \bar{\partial} \hat{r})$ is a total derivative so does not contribute to the classical equation of motion and we will drop it from the discussion here. An analog of (\ref{poincare_1st}) is obtained (at large $k$) as follows
\begin{equation}\label{action}
	\begin{aligned}
		& S_{cl}=\frac{k}{2 \pi}\int 2 d^2 \sigma \(\partial \hat{r} \bar{\partial} \hat{r}+\partial \hat{\xi} \bar{\partial} \hat{\xi} +  \hat{\bar{\chi}}\partial (\hat{\xi} -i \hat{\theta})+\hat{\chi} \bar{\partial} (\hat{\xi} +i \hat{\theta})-\frac{1}{\sinh^2  \hat{r}}\hat{\chi}\hat{\bar{\chi}} \)\\
	\end{aligned}
\end{equation}
We define new sets of coordinates 
\begin{equation}
	\begin{aligned}
		& r:=\alpha_+ \hat{r},\\& \chi:=-\frac{\alpha_+}{\sqrt{2}}\(\hat{\chi}+\frac{1}{2}\partial \hat{\xi} \), ~~ \bar{\chi}:=-\frac{\alpha_+}{\sqrt{2}}\(\hat{\bar{\chi}}+\frac{1}{2}\bar{\partial} \hat{\xi} \),\\& W:=\frac{\alpha_+}{\sqrt{2}}\(\hat{\xi}+i \hat{\theta}\), ~~~~~~ \bar{W}:=\frac{\alpha_+}{\sqrt{2}}\(\hat{\xi}-i \hat{\theta}\)
	\end{aligned}
\end{equation}

Near the boundary of $AdS$ the action (\ref{action}) reduces to the following free theory (valid for all $k$) 
\begin{equation}\label{free}
	\begin{aligned}
		& S=\frac{1}{ \pi}\int  d^2 \sigma \(\frac{1}{2}\partial r \bar{\partial} r-\frac{1}{4 \alpha_+}r R -\chi \bar{\partial} W -\bar{\chi} \partial \bar{W} \)\\
	\end{aligned}
\end{equation}
Conserved currents (all values of $k$) are\footnote{At large $k$ expression for the currents follow from a change of variables.}
\begin{equation}
	\begin{aligned}
		& T=-\frac{1}{2} (\partial r)^2-\frac{1}{\alpha_+}\partial^2 r+\chi \partial W\\
		& J^3=\sqrt{k}\(\chi -\frac{1}{4}\partial W\)
	\end{aligned}
\end{equation}
The vertex operators in this theory (\ref{free}) can be labeled as ($\nu$ is the winding around the $\hat{\theta}$ circle) 
\begin{equation}\label{vertex}
	\begin{aligned}
		& \tilde{V}^\nu_{j,m,\bar{m}}=N^\nu_{j,m,\bar{m}} \ e^{(m-\frac{k \nu}{4})\frac{W}{\sqrt{k}}+(\bar{m}-\frac{k \nu}{4})\frac{\bar{W}}{\sqrt{k}}} \ e^{\sqrt{k}\nu \int_{(0,0)}^{(z,\bar{z})}\chi dz'+\sqrt{k}\nu \int_{(0,0)}^{(z,\bar{z})}\bar{\chi} d\bar{z}'} \ e^{\frac{2}{\alpha_+}jr}
	\end{aligned}
\end{equation}
The normalization is to be fixed later.

The OPE with $T$ gives
\begin{equation}
	\begin{aligned}
		&h'=-\frac{j(j+1)}{k-2}+\nu \(m-\frac{k\nu}{4}\)
	\end{aligned}
\end{equation}
The OPE with $J^3$ gives
\begin{equation}
	\begin{aligned}
		&m'=m-\frac{k\nu}{2}
	\end{aligned}
\end{equation}
We consider operators that are uncharged under $J^3,\bar{J}^3$ (these are the operators that are unmodified if we gauge the translation generated by $J^3,\bar{J}^3$), i.e., $$m=\bar{m}=\frac{k\nu}{2}$$
Among these, marginal ones satisfy
\begin{equation}
	\begin{aligned}
		& 1=-\frac{j(j+1)}{k-2}+\frac{k\nu^2}{4}
	\end{aligned}
\end{equation}
If we choose $\nu=\pm1$, we get $$j=-\frac{1}{2}(1\pm(k-3)) $$ The operator that decays at large $r$ is given by the choice of $+$ sign above (notice that the vertex operators don't decay at the boundary).

We consider the action below as the definition of a new conformal field theory
\begin{equation}\label{winding}
	\begin{aligned}
		& \tilde{S}=\frac{1}{ \pi}\int  d^2 \sigma \(\frac{1}{2}\partial r \bar{\partial} r-\frac{1}{4 \alpha_+}r R -\chi \bar{\partial} W -\bar{\chi} \partial \bar{W} +\frac{\pi\mu}{\alpha_+^2} (V^{+}+ V^{-})\)\\
	\end{aligned}
\end{equation}
Where the interaction term is now modified to
\begin{equation}\label{winding_potential}
	\begin{aligned}
				& V^{\pm}=\ e^{\pm \frac{\sqrt{k}}{4}(W+\bar{W})} \ e^{\pm \sqrt{k} (\int_{(0,0)}^{(z,\bar{z})}\chi dz'+ \int_{(0,0)}^{(z,\bar{z})}\bar{\chi} d\bar{z}')} \ e^{-\frac{\alpha_+}{2}r}\propto  \tilde{V}^{\pm1}_{-(k-2)/2,\pm k/2,\pm k/2}
	\end{aligned}
\end{equation}

 We would like to save the notation $r, \chi, W,\bar{W}$ for the co-ordinates in the global patch of AdS and change our notation for the new conformal theory (\ref{winding}) to the following 
\begin{equation}\label{winding_potential}
	\begin{aligned}
				& \beta:=-\sqrt{k}\ \chi, ~~ \gamma:=\frac{1}{\sqrt{k}}\ W, ~~ \varphi:=-\frac{1}{\sqrt{2}}r
	\end{aligned}
\end{equation}
The action takes the following form \cite{Jafferis:2021ywg}
\begin{subequations}
\begin{empheq}{align}\label{winding2}
  & \tilde{S}=\frac{1}{ \pi}\int  d^2 \sigma (\partial \varphi \bar{\partial} \varphi+\frac{Q}{4}\varphi R +\beta \bar{\partial} \gamma +\bar{\beta} \partial \bar{\gamma} +\frac{\pi\mu}{\alpha_+^2} (V^{+}+ V^{-}))\\ 
  & V^{\pm}:=\ e^{\pm \frac{k}{4}(\gamma+\bar{\gamma})} \ e^{\mp  (\int_{(0,0)}^{(z,\bar{z})}\beta dz'+ \int_{(0,0)}^{(z,\bar{z})}\bar{\beta} d\bar{z}')} \ e^{2b\varphi}, ~ Q:=\frac{1}{2b}:=\frac{1}{\sqrt{k-2}}
\end{empheq}
\end{subequations}
The measure for path integration is the usual Liouville measure for $\varphi$ (the precise normalization of the zero mode integration is given in (\ref{map})) and the standard measure for $\beta\gamma$ system.
The conserved currents are
\begin{equation}
	\begin{aligned}
		& T=- (\partial \varphi)^2+Q\partial^2 \varphi-\beta \partial \gamma\\
		& J^3=-\beta -\frac{k}{4}\partial \gamma
	\end{aligned}
\end{equation}
Relevant OPE are
\begin{equation}\label{OPE}
	\begin{aligned}
		\beta(z)\gamma(0)\sim -\frac{1}{z}, ~~~ \varphi(z, \bar{z})\varphi(0,0)\sim -\frac{1}{2}(\ln z+\ln \bar{z})
	\end{aligned}
\end{equation}
Vertex operators are
\begin{equation}\label{vo}
	\begin{aligned}
		\tilde{V}^\nu_{j,m,\bar{m}}=N^\nu_{j,m,\bar{m}} \ e^{a \gamma+\bar{a} \gamma} \ e^{n(\int_{(0,0)}^{(z,\bar{z})}\beta dz'+ \int_{(0,0)}^{(z,\bar{z})}\bar{\beta} d\bar{z}')} \ e^{2\alpha \varphi}
	\end{aligned}
\end{equation}
with the following map
\begin{equation}
	\begin{aligned}
		& \alpha:=-\frac{j}{2b},~~ 
		& a:=m-\frac{k\nu}{4}, ~~
		& n:=-\nu
	\end{aligned}
\end{equation}

\section{Proving the equivalence}\label{Proving the equivalence}
\textbf{Statement of the duality}: Correlation function of delta function normalizable states of the worldsheet CFT are mapped as $\tilde{V}^0_{j,m,\bar{m}} \leftrightarrow V^0_{j,m,\bar{m}}$ under the mapping of parameters given by\footnote{The calculation presented in this note can be extended to the spectrally flowed vertex operators presented in \ref{vo}, however the one has to be very careful about the normalization and the OPE. We will leave that to future work.}
\begin{subequations}
\begin{empheq}{align}\label{duality_map}
	& N^{0}_{j_a,m_a,\bar{m}_a}:=\frac{1}{\pi}, ~~b:=\frac{b'}{2}, ~	\mu:=b'^2 \alpha_+^2\(\frac{\mu'\gamma(b'^2)}{\pi }\)^{1/2}\\
	 & b':=\frac{1}{b''}, ~~ \pi \mu'\gamma(b'^2):=(\pi \mu''\gamma(b''^2))^{1/(b''^2)}\\
	 & b''=\frac{1}{\sqrt{k-2}}, ~~ \mu''=\frac{b''^2}{\pi^2}
\end{empheq}
\end{subequations}
The first of these equations defines the Liouville parameter $(b',\mu')$ which is then related to $(b'',\mu'')$ through the Liouville duality. The last of these equations is a well-known map relating sigma model correlation functions and those of a Liouville theory as we review next. 

\subsection{Sigma model correlation functions}\label{SMcf}

The correlation functions (of normalizable operators, to be specific) in the $H_3^+$ sigma model can be written in terms of a  corresponding correlation function in Liouville's theory. This has been proven in \cite{Ribault:2005wp} using the fact that both sides satisfy the same Knizhnik-Zamolodchikov (KZ) equations, given that the solution of the KZ equation is uniquely specified by its behavior in the OPE limit, which leads to a recursion relation. In this subsection, we review the results in brief.
Consider a Liouville theory (see \ref{LT} for conventions) parameters
\begin{equation}
	b'':=\frac{1}{\sqrt{k-2}}, ~~ \mu'':=\frac{b''^2}{\pi^2}
\end{equation}
With each vertex operator in the sigma model description, we associate an operator in Liouville theory with the map
\begin{equation}
	\alpha'':=b''\(j''+1+\frac{1}{2b''^2}\), ~~ j'':=-(j+1)
\end{equation}
Before proceeding forward we pause for a few comments regarding this assignment. Note that the large $k$ limit maps to the classical limit of the Liouville theory, however, the corresponding operators at fixed $j$ become heavy and the discussion is still somewhat involved. 

The two-point function can be written as (here $R$ is a large positive number and the formulas are presented with a suitable divergent power of $R$ stripped off associated to the dimension of the external operator)
\begin{equation}\label{sg2}
	\begin{aligned}
		&~~ \langle  V^{0}_{j_1,m_1,\bar{m}_1}(0,0) V^{0}_{j_2,m_2,\bar{m}_2}(R,R)  \rangle \\
		& ~~~~~~~~~~~~~~~~~~=2 \pi b'' \( \prod_{a=1,2}\frac{1}{\pi} \frac{\Gamma(-j''_a+m_a)}{\Gamma(1+j''_a-\bar{m}_a)}\) 
	\langle V_{\alpha''_1}(0,0)V_{\alpha''_2}(R,R) \rangle_{(b'',\mu'')} 
	\end{aligned}
\end{equation}
The three-point function can be written as 
\begin{equation}\label{sg3}
	\begin{aligned}
		&~~ \ \langle  V^{0}_{j_1,m_1,\bar{m}_1}(0,0)  V^{0}_{j_2,m_2,\bar{m}_2}(1,1)  V^{0}_{j_3,m_3,\bar{m}_2}(R,R)  \rangle\\
		&=2\pi b''\( \prod_{a=1,2,3} \frac{1}{\pi} \frac{\Gamma(-j''_a+m_a)}{\Gamma(1+j''_a-\bar{m}_a)}\)\\& 
		~~~\int  \frac{d^2z}{\pi} \[ z^{-(k/2+m_1)}\bar{z}^{-(k/2+\bar{m}_1)} (1-z)^{-(k/2+m_2)}(1-\bar{z})^{-(k/2+\bar{m}_2)}    \] \\&~~~~~~~~~~~~~~~~~~~~~~~~\langle V_{\alpha''_1}(0,0)V_{\alpha''_2}(1,1)V_{\alpha''_3}(R,R)V_{-\frac{1}{2b''}}(z,\bar{z}) \rangle_{(b'',\mu'')}
	\end{aligned}
\end{equation}
Both of them are subject to the condition that 
\begin{equation}
	\sum_a m_a=\sum_a \bar{m}_a=0
\end{equation}
The correlation functions $\langle ..\rangle_{b,\mu}$ are in a Liouville theory with corresponding parameters (for an explicit expression and modern discussion of the correlation functions including winding sectors, see \cite{Dei:2021xgh, Dei:2021yom}).

\subsection{Winding condensate correlation functions}

We would like to compute the following correlation function in the worldsheet CFT given by (\ref{winding2}) with the usual measure on the Liouville field $\varphi$ and $\beta-\gamma$ system
\begin{equation}
	\begin{aligned}
		\langle \prod_{a} \tilde{V}^{\nu_a}_{j_a,m_a,\bar{m}_a}(w_a,\bar{w}_a)  \rangle
	\end{aligned}
\end{equation}
We will restrict ourselves to sphere topology.
To this end, we perform the path integral over the Liouville zero mode $\varphi_0$ defined by ($\varphi_0$ denotes the kernel of the scalar Laplacian and $\varphi'(z,\bar{z})$ is the space of functions orthogonal to the kernel) \cite{PhysRevLett.66.2051, Dorn:1994xn, Zamolodchikov:1995aa, Teschner:2001rv}
\begin{equation}
	\varphi(z,\bar{z})=\varphi_0+\varphi'(z,\bar{z})
\end{equation}
using the following identity
\begin{equation}
	\begin{aligned}
	\int_{-\infty}^{+\infty}d\varphi_0 \ e^{2a\varphi_0-\alpha e^{2b\varphi_0}}=\frac{1}{2b}\Gamma\(\frac{a}{b}\)\alpha^{-\frac{a}{b}}
	\end{aligned}
\end{equation}
We are left with the following integral over $\varphi'$ alone
\begin{equation}\label{cor1}
	\begin{aligned}
		& ~~\langle \prod_{a} \tilde{V}^{\nu_a}_{j_a,m_a,\bar{m}_a}(w_a,\bar{w}_a)  \rangle\\
		=& ~~\frac{\mathcal{N}}{2b}\Gamma\(-s\)\frac{\mu^s}{\alpha_+^{2s}}\langle \prod_{a} \tilde{V}^{\nu_a}_{j_a,m_a,\bar{m}_a} (w_a,\bar{w}_a) (\int  d^2 z \ (V^{+}(z,\bar{z})+ V^{-}(z,\bar{z})))^{s}\rangle'
	\end{aligned}
\end{equation}
Here $\mathcal{N}$ is an arbitrary normalization factor associated with the correct measure for the $\varphi_0$ path integral that will be determined below. We have defined the following Liouville-like parameters 
\begin{equation}\label{ac}
	s=\frac{Q-\sum_a \alpha_a}{b} =\frac{1+\sum_aj_a}{2b^2}
	\end{equation}
Here $\langle .. \rangle'$ denotes path integration only over $\varphi'$ with  the action $\tilde{S}$ under replacement $\varphi\to\varphi'$ and $V^\pm \to 0$. Therefore for the path integration only over $\varphi'$ we have additional conservation laws that we mention below

\textbf{Winding conservation:}
The only contribution comes from the $(V^+)^{s'+n}(V^-)^{s'}$ term, where 
\begin{equation}\label{wc}
	(s'+n)-s'-\sum_an_a=0 \implies n=\sum_a n_a
\end{equation}
We will assume $s'=(s-n)/2$ is an integer for the argument presented here and define the answer by analytic continuation to the physical, normalizable values. Because of the factor of $\Gamma(-s)$ in (\ref{cor1}) we get a pole for positive integer values of $s=2s'+n$. We will be interested in the residue at these poles.

Further note that both $n,s'$ are determined in terms of external quantum numbers. This is a huge improvement because the sum of Feynman diagrams is reduced to a single term.

\textbf{$J^3$ conservation:}
This gives us the following selection rule
\begin{equation}\label{ac2}
	\begin{aligned}
		\sum_a m_a+\frac{k}{2}(s'+n-s')=0 \implies \sum_c a_c+\frac{k}{4}\sum_c n_c=0
	\end{aligned}
\end{equation}
There is a similar condition for $\bar{a}$.

Now we turn to (\ref{cor1}) and perform Wick's contractions using (\ref{OPE}) to obtain\footnote{We are using $A_{(a}B_{b)}=A_aB_b+A_bB_a$}
\begin{equation}\label{Wick}
	\begin{aligned}
		&~~\res_{s\to 2s'+n} \  \langle \prod_{a} \tilde{V}^{\nu_a}_{j_a,m_a,\bar{m}_a}(w_a,\bar{w}_a)  \rangle\\
		&=\frac{\mathcal{N}}{2b}\frac{(-1)^s}{\Gamma(s+1)} \frac{\mu^s}{\alpha_+^{2s}}\frac{\Gamma(s+1)}{\Gamma(s-s'+1)\Gamma(s'+1)}\prod_a N^{\nu_a}_{j_a,m_a,\bar{m}_a} \\
		&~~~~\prod_{a<b} \[ w_{ab}^{-2\alpha_a\alpha_b-a_{(a}n_{b)}}  \bar{w}_{ab}^{-2\alpha_a\alpha_b-\bar{a}_{(a}n_{b)}} \] \int\prod_{i=1}^{s'}  d^2z_i \int \prod_{i'=1}^{s'}  d^2z'_{i'} \\
		&~~~~\prod_{i=1}^{s'+n}\prod_{a} \[ (z_i-w_a)^{-2\alpha_a b+a_a-\frac{k}{4}n_{a}}  (\bar{z}_i-\bar{w}_a)^{-2\alpha_a b +\bar{a}_a-\frac{k}{4}n_{a}}\]\\
		&~~~~\prod_{i'=1}^{s'}  \prod_{a} \[ (z'_{i'}-w_a)^{-2\alpha_a b-a_a+\frac{k}{4}n_{a}}  (\bar{z}'_{i'}-\bar{w}_a)^{-2\alpha_a b -\bar{a}_a+\frac{k}{4}n_{a}} \]\\
		&~~~~\prod_{i'<j'}  \[ (z_{i'j'})^{\frac{k}{2}-2b^2}  (\bar{z}_{i'j'})^{\frac{k}{2}-2b^2}\] \ \prod_{i<j}   \[ (z_{ij})^{\frac{k}{2}-2b^2}  (\bar{z}_{ij})^{\frac{k}{2}-2b^2}\]\\
		&~~~~ \prod_{i,j'} \[ (z_{ij'})^{-\frac{k}{2}-2b^2}  (\bar{z}_{ij'})^{-\frac{k}{2}-2b^2}\]
	\end{aligned}
\end{equation}
We have defined
\begin{equation}
	w_{ab}=w_a-w_b,~~z_{ij}=z_i-z_j,~~z_{i'j'}=z_{i'}-z_{j'},~~z_{ij'}=z_{i}-z_{j'}
\end{equation}
The first line after the numerical factor\footnote{Coming from combinatorial factor} is obtained from contraction between two external vertex operators, second line from contraction between one external vertex operator and $V^+$, third line from contraction between one external vertex operator and $V^-$, fourth line from $V^--V^-$,$V^+-V^+$ contraction, and the final line is from $V^+-V^-$ contraction. This can be simplified by writing it in terms of worldsheet $SL(2,R)$ charges as follows. 
\begin{equation}
	-2\alpha_a b=j_a, ~~a_a-\frac{k}{4}n_a=m_a, ~~\bar{a}_a-\frac{k}{4}n_a=\bar{m}_a
\end{equation}
The result is
\begin{equation}\label{res}
	\begin{aligned}
		&~~\res_{s\to 2s'+n} \ \langle \prod_{a} \tilde{V}^{\nu_a}_{j_a,m_a,\bar{m}_a}(w_a,\bar{w}_a)  \rangle\\
		&=\frac{\mathcal{N}}{2b}\frac{(-\mu)^s}{\alpha_+^{2s}}\frac{1}{\Gamma(s-s'+1)\Gamma(s'+1)}\prod_a N^{\nu_a}_{j_a,m_a,\bar{m}_a}\\
		&~~~~\prod_{a<b} \[ w_{ab}^{-2\alpha_a\alpha_b-a_{(a}n_{b)}}  \bar{w}_{ab}^{-2\alpha_a\alpha_b-\bar{a}_{(a}n_{b)}} \] \int\prod_{i=1}^{s'}  d^2z_i \int \prod_{i'=1}^{s'}  d^2z'_{i'} \\
		&~~~~\prod_{i=1}^{s'+n}  \prod_{a} \[ (z_i-w_a)^{j_a+m_a}  (\bar{z}_i-\bar{w}_a)^{j_a+\bar{m}_a}\]\\
		&~~~~\prod_{i'=1}^{s'}  \prod_{a} \[ (z'_{i'}-w_a)^{j_a-m_a}  (\bar{z}'_{i'}-\bar{w}_a)^{j_a-\bar{m}_a} \]\\
		&~~~~\prod_{i'<j'}   \[ (z_{i'j'})  (\bar{z}_{i'j'})\] \ \prod_{i<j}   \[ (z_{ij})  (\bar{z}_{ij})\]\\
		&~~~~ \prod_{i,j'}   \[ (z_{ij'})^{1-k}  (\bar{z}_{ij'})^{1-k}\]
	\end{aligned}
\end{equation}
From now on, for the purpose of proving the duality we will restrict to $H_3^+$ model, i.e., consider external operators satisfying
\begin{equation}\label{H3+}
	n_a=0 \implies s'=\frac{s}{2}
\end{equation}
This in particular implies
\begin{equation}\label{H3+2}
	s'=\frac{s}{2},~~ \sum_am_a=0, ,~~ \sum_a \bar{m}_a=0
\end{equation}

We would like to define the value of the correlation function in the complex plane through analytic continuation. However, the free field-like calculation presented here allows us to look at only values of the parameter $s$ where we have a pole due to the presence of the factor $\Gamma(-s)$\footnote{ Similar comment applies for the Liouville theory itself, where we can only recover half the poles using this method.}. We will focus on these poles only. Note that when $s$ is an odd integer, the residue on the pole vanishes due to winding conservation - in other words, there is no pole at these values.

\subsubsection{Two point function}
We restrict to $m_a=\bar{m}_a$ for all the external operators.\footnote{ We restrict to this special sector due to the technical difficulty of generalizing the identity presented in  \ref{identity}. }  Set $w_1=0$, and $w_2=R\to \infty$ above and strip off the divergent power of $R$ (which determines the worldsheet conformal dimension of the operator\footnote{For two-point function of two operators with $\alpha_1=\alpha_2=\alpha, n_1=-n_2=n, a_1=-a_2=a$ the stripped off factor is $R^{-4(\alpha^2-j(1+2j)Q^2-an) }=R^{-4(-Q^2j(j+1)-an) }$.}) to give
\begin{equation}
	\begin{aligned}
		&~~\res_{s\to 2s'} \ \langle  \tilde{V}^{0}_{j_1,m_1,\bar{m}_1}(0,0) \tilde{V}^{0}_{j_2,m_2,\bar{m}_2}(R,R)  \rangle\\
		&=\frac{\mathcal{N}}{2b}\frac{(-\mu)^s}{\alpha_+^{2s}}\frac{1}{\Gamma(s-s'+1)\Gamma(s'+1)} \prod_{a=1,2} N^{0}_{j_a,m_a,\bar{m}_a} \int\prod_{i=1}^{s'}  d^2z_i \int \prod_{i'=1}^{s'}  d^2z'_{i'}\\
		&~~~~\prod_{i=1}^{s'}  \[ |z_i|^{2j_1+2m_1} \]
		~~\prod_{i'=1}^{s'}  \[ |z'_{i'}|^{2j_1-2m_1}  \]~~\prod_{i'<j'}  \[ |z_{i'j'}|^2  \] ~~\prod_{i<j}   \[ |z_{ij}|^2 \]\ ~~  \prod_{i,j'}   \[ |z_{ij'}|^{2-2k}  \]
	\end{aligned}
\end{equation}
Perform the $z_i$ integrations first using \ref{identity}. This simplifies considerably because the $z_i$ integral does not produce any dual integral.\footnote{We used the fact that $s$ is an even integer.}
\begin{equation}
	\begin{aligned}
		&~~\res_{s\to 2s'} \ \langle  \tilde{V}^{0}_{j_1,m_1,\bar{m}_1}(0,0) \tilde{V}^{0}_{j_2,m_2,\bar{m}_2}(R,R)  \rangle\\
		&=\frac{\mathcal{N}}{2b}\frac{\mu^s}{\alpha_+^{2s}}\frac{1}{\Gamma(s-s'+1)\Gamma(s'+1)}
		\prod_{a=1,2} N^{0}_{j_a,m_a,\bar{m}_a}\\
		&~~~~
		~~~~ \int d^2z'_{i'}\prod_{i'=1}^{s'} ~~\prod_{i'=1}^{s'}\[ |z'_{i'}|^{2j_1-2m_1}  \]~~\prod_{i'<j'}   \[ |z_{i'j'}|^2  \] 		M(0,s',\{t_j\},\{p_j\})
	\end{aligned}
\end{equation}
Here we have defined 
\begin{equation}
	M(m,n,\{t_j\},\{p_j\})=\frac{\pi^n \Gamma(n+1)}{\pi^m \Gamma(m+1)}\frac{\prod_{j=1}^{n+m+1}\gamma(1+p_j)}{\gamma(1+n+\sum_{j=1}^{n+m+1} p_j)}\prod_{1\leq j<j'\leq n+m+1}|t_j-t_{j'}|^{2p_j+2p_{j'}+2}
\end{equation}
and
\begin{equation}
	 t_{j}=z'_{j},~p_j=1-k, ~ t_{s'+1}=0,~p_{s'+1}=j_1+m_1
\end{equation}
The above expression can be simplified to
\begin{equation}
	\begin{aligned}
		&~~\res_{s\to 2s'+n} \ \langle  \tilde{V}^{0}_{j_1,m_1,\bar{m}_1}(0,0) \tilde{V}^{0}_{j_2,m_2,\bar{m}_2}(R,R)  \rangle\\
		&=\frac{\mathcal{N}}{2b}\frac{\mu^s}{\alpha_+^{2s}}\frac{1}{\Gamma(s-s'+1)\Gamma(s'+1)}
		\prod_{a=1,2} N^{0}_{j_a,m_a,\bar{m}_a}\\
		& ~~~~~ \pi^{s'} \Gamma(s'+1)  \gamma(2-k)^{s'} \gamma(1+j_1+m_1) \gamma(1+j_2+m_2)  \\
		&~~~~
		~~~~\int\prod_{i'=1}^{s'}  d^2z'_{i'} ~~\prod_{i'=1}^{s'}\[ |z'_{i'}|^{4j_1+2(2-k)}  \]~~\prod_{i'<j'}   \[ |z_{i'j'}|^{4(2-k)}  \] 		
	\end{aligned}
\end{equation}
We use the following formula (using (\ref{ac}), (\ref{ac2}))
\begin{equation}
	\gamma(1+s'(2-k)+j_1+m_1)=\gamma(-j_2-m_2)=\frac{1}{\gamma(1+j_2+m_2)}
\end{equation}
The above answer can be written in terms of a Liouville correlation function where the Liouville theory (see Appendix \ref{LT}) is defined by $b'=\sqrt{k-2}, ~ \mu'$ (to be expressed in terms of $\mu$ below).
With each external vertex operator, we associate a Liouville operator by
\begin{equation}
	~\alpha'_a=\frac{1}{b'}\(\frac{k-2}{2}-j_a \) \implies \sum_{a=1,2} \alpha'_a = (Q'-s'b')  
\end{equation}
The final answer is
\begin{equation}
	\begin{aligned}
		&~~\res_{s\to 2s'} \ \langle  \tilde{V}^{0}_{j_1,m_1,\bar{m}_1}(0,0) \tilde{V}^{0}_{j_2,m_2,\bar{m}_2}(R,R)  \rangle\\
		&=\frac{\mathcal{N}}{2b}\frac{\mu^s\pi^{s/2}}{\alpha_+^{2s}}\frac{1}{\Gamma(s/2+1)}\(\frac{-1}{b'^4\gamma(b'^2)}\)^{s/2} \( \prod_a \gamma(1+j_a+m_a)\) \\
		& ~~~~~      \prod_{a=1,2} N^{0}_{j_a,m_a,\bar{m}_a} ~~\int\prod_{i'=1}^{s'}  d^2z'_{i'} ~~\prod_{i'=1}^{s'}\[ |z'_{i'}|^{-4b'\alpha'_1}  \]~~\prod_{i'<j'}   \[ |z_{i'j'}|^{-4b'^2}  \] 	\\
		&=\frac{\mathcal{N}}{2b}\frac{\mu^s\pi^{s/2}}{\alpha_+^{2s}b'^{2s}}\frac{1}{\gamma(b'^2)^{s/2}\mu'^{s/2}} \( \prod_{a=1,2} \gamma(1+j_a+m_a)\) \prod_{a=1,2} N^{0}_{j_a,m_a,\bar{m}_a}\\
		& ~~~~~~~~~~~~~~~~~~~~~~\lim_{\alpha'_3\to 0}~~ \res_{\sum_{a=1}^{3} \alpha'_a \to (Q'-s'b')  }\langle V_{\alpha'_1}(0,0)V_{\alpha'_3}(1,1)V_{\alpha'_2}(R,R) \rangle_{(b',\mu')}
	\end{aligned}
\end{equation}
In the first line we used \ref{property}. To get to the second line from the first we used \ref{LT1}.
If we take
\begin{equation}\label{map}
	\mu=b'^2 \alpha_+^2\(\frac{\mu'\gamma(b'^2)}{\pi }\)^{1/2}, ~~ N^{0}_{j_a,m_a,\bar{m}_a}=\frac{1}{\pi},~~ \mathcal{N}=2\pi
\end{equation}
this precisely matches with the corresponding residues that can be obtained from (\ref{sg2}) if we identify Liouville theory with $(b',\mu')$ (after exchanging the limit and the residue operation) with the dual Liouville theory with $(b'',\mu'')$.

The above calculation goes through essentially unchanged if we consider spectrally flowed vertex operators such that the sum over winding is zero (suitable factors of $R$ are scaled away and again we are considering only $m_a=\bar{m}_a$)
\begin{equation}
	\begin{aligned}
		&~~ \ \langle  \tilde{V}^{+\nu}_{j_1,m_1,\bar{m}_1}(0,0) \tilde{V}^{-\nu}_{j_2,m_2,\bar{m}_2}(R,R)  \rangle\\
		&=\frac{2\pi}{ b'} \( \prod_{a=1,2} \gamma(1+j_a+m_a)\) \prod_{a=1,2} N^{0}_{j_a,m_a,\bar{m}_a}\\
		& ~~~~~~~~~~~~~~~~~~~~~~\lim_{\alpha'_3\to 0}~~ \langle V_{\alpha'_1}(0,0)V_{\alpha'_3}(1,1)V_{\alpha'_2}(R,R) \rangle_L  \\
	\end{aligned}
\end{equation}
If we apply this formula for the winding potential $V^\pm$ given in \ref{winding2} the factor $$ \prod_{a=1,2} \gamma(1+j_a+m_a)$$ diverges for integer $k>3$, which is correlated with the fact that $\mu^2 \sim \gamma(k-2)$ has a zero for integer $k>3$.  As a result, if we consider two-point function of $\mu V^{\pm}$, then we get a finite answer. This explains why the deformation of the free theory by $(\mu/\alpha^2_+) (V^{+}+V^{-})$ considered in this paper is a sensible one.

\subsubsection{Three point function}

We restrict to $m_a=\bar{m}_a$ for all the external operators. Set $w_1=0, w_2=1$, and $w_3=R\to \infty$ above and strip off the divergent power of $R$ (this determines the worldsheet conformal dimension of the operator) to give
\begin{equation}
	\begin{aligned}
		&~~\res_{s\to 2s'} \ \langle  \tilde{V}^{0}_{j_1,m_1,\bar{m}_1}(0,0)  \tilde{V}^{0}_{j_2,m_2,\bar{m}_2}(1,1)  \tilde{V}^{0}_{j_3,m_3,\bar{m}_2}(R,R)  \rangle\\
		&=\frac{\mathcal{N}}{2b}\frac{(-\mu)^s}{\alpha_+^{2s}}\frac{1}{\Gamma(s-s'+1)\Gamma(s'+1)} \prod_{a=1}^{3} N^{0}_{j_a,m_a,\bar{m}_a} \int\prod_{i=1}^{s'}  d^2z_i \int \prod_{i'=1}^{s'}  d^2z'_{i'}\\
		&~~~~\prod_{i=1}^{s'}  \[ |z_i|^{2j_1+2m_1}|1-z_i|^{2j_2+2m_2} \]
		~~\prod_{i'=1}^{s'}  \[ |z'_{i'}|^{2j_1-2m_1} |1-z'_{i'}|^{2j_2-2m_2}  \] \\
		&~~\prod_{i'<j'}  \[ |z_{i'j'}|^2  \] ~~\prod_{i<j}   \[ |z_{ij}|^2 \]\ ~~  \prod_{i,j'}   \[ |z_{ij'}|^{2-2k}  \]
	\end{aligned}
\end{equation}
Perform the $z_i$ integrations first using \ref{identity}. This gives only one dual integral
\begin{equation}
	\begin{aligned}
		&~~\res_{s\to 2s'} \ \langle  \tilde{V}^{0}_{j_1,m_1,\bar{m}_1}(0,0)  \tilde{V}^{0}_{j_2,m_2,\bar{m}_2}(1,1)  \tilde{V}^{0}_{j_3,m_3,\bar{m}_2}(R,R)  \rangle\\
		&=\frac{\mathcal{N}}{2b}\frac{\mu^s}{\alpha_+^{2s}}\frac{1}{\Gamma(s-s'+1)\Gamma(s'+1)}
		 \prod_{a=1}^{3} N^{0}_{j_a,m_a,\bar{m}_a}\\
		&~~~~
		~~~~ \int d^2z'_{i'}\prod_{i'=1}^{s'} ~~\prod_{i'=1}^{s'}\[ |z'_{i'}|^{2j_1-2m_1}|1-z'_{i'}|^{2j_2-2m_2}  \]~~\prod_{i'<j'}   \[ |z_{i'j'}|^2  \] 		M(1,s',\{t_j\},\{p_j\})
	\end{aligned}
\end{equation}
Here
\begin{equation}
	 t_{j}=z'_{j},~p_j=1-k, ~ t_{s'+1}=0,~p_{s'+1}=j_1+m_1,  ~ t_{s'+2}=1,~p_{s'+2}=j_2+m_2
\end{equation}
The above expression can be simplified to
\begin{equation}
	\begin{aligned}
		&~~\res_{s\to 2s'} \ \langle  \tilde{V}^{0}_{j_1,m_1,\bar{m}_1}(0,0)  \tilde{V}^{0}_{j_2,m_2,\bar{m}_2}(1,1)  \tilde{V}^{0}_{j_3,m_3,\bar{m}_2}(R,R)  \rangle\\
		&=\frac{\mathcal{N}}{2b}\frac{\mu^s}{\alpha_+^{2s}}\frac{1}{\Gamma(s-s'+1)\Gamma(s'+1)}
		 \prod_{a=1}^{3} N^{0}_{j_a,m_a,\bar{m}_a}\\
		& ~~~~~~~~~~ \pi^{s'-1} \Gamma(s'+1)  \gamma(2-k)^{s'} \gamma(1+j_1+m_1) \gamma(1+j_2+m_2) \gamma(1+j_3+m_3)  \\
		& 		~~~~~~~~\int  d^2z \[ |z|^{-2(j_1+m_1+1)} |1-z|^{-2(j_2+m_2+1)}    \] \\
		&~~~~
		~~~~\int\prod_{i'=1}^{s'}  d^2z'_{i'} ~~\prod_{i'=1}^{s'}\[ |z'_{i'}|^{4j_1+2(2-k)} |1-z'_{i'}|^{4j_2+2(2-k)} |z'_{i'}-z|^{-2(2-k)} \]~~\prod_{i'<j'}   \[ |z_{i'j'}|^{4(2-k)}  \] 	\\
	\end{aligned}
\end{equation}

The above answer can be written in terms of a Liouville correlation function where the Liouville theory (see \ref{LT}) is defined by $b'=\sqrt{k-2}, ~ \mu'$ (to be related to $\mu$ in (\ref{dualitymap})).
With each external vertex operator, we associate a Liouville operator by
\begin{equation}
	~\alpha'_a=\frac{1}{b'}\(\frac{k-2}{2}-j_a \) \implies \sum_{a=1,2,3} \alpha'_a = Q'-s'b'+\frac{b'}{2}
\end{equation}
The final answer is
\begin{equation}
	\begin{aligned}
		&~~\res_{s\to 2s'} \ \langle  \tilde{V}^{0}_{j_1,m_1,\bar{m}_1}(0,0)  \tilde{V}^{0}_{j_2,m_2,\bar{m}_2}(1,1)  \tilde{V}^{0}_{j_3,m_3,\bar{m}_2}(R,R)  \rangle\\
		&=\frac{\mathcal{N}}{2b}\frac{\mu^s\pi^{s/2-1}}{\alpha_+^{2s}}\frac{1}{\Gamma(s/2+1)}\gamma(-b'^2)^{s/2} \( \prod_{a=1}^{3} \gamma(1+j_a+m_a)\) ~~ \prod_{a=1}^{3} N^{0}_{j_a,m_a,\bar{m}_a} \\
		& ~~~\int  d^2z \[ |z|^{-2(j_1+m_1+1)} |1-z|^{-2(j_2+m_2+1)}    \] \\
		&~~~\int\prod_{i'=1}^{s'}  d^2z'_{i'} ~~\prod_{i'=1}^{s'}\[ |z'_{i'}|^{-4b'\alpha'_1} |1-z'_{i'}|^{-4b'\alpha'_2}|z'_{i'}-z|^{2b'^2}|  \]~~\prod_{i'<j'}   \[ |z_{i'j'}|^{-4b'^2}  \] 	\\
		&=\frac{1}{2b}\frac{\mu^s\pi^{s/2}}{\alpha_+^{2s}b'^{2s}}\frac{1}{\gamma(b'^2)^{s/2}\mu'^{s/2}} \( \prod_{a=1}^{3} \gamma(1+j_a+m_a)\)\prod_{a=1}^{3} N^{0}_{j_a,m_a,\bar{m}_a}\\
		& ~~~\int  \frac{d^2z}{\pi} \[ |z|^{-(k+2m_1)} |1-z|^{-(k+2m_2)}    \] \res_{\sum_{a=1}^{3} \alpha'_a \to (Q'-s'b'+\frac{b'}{2})  }\langle V_{\alpha'_1}(0,0)V_{\alpha'_2}(1,1)V_{\alpha'_3}(R,R)V_{-\frac{b'}{2}}(z,\bar{z}) \rangle_{(b',\mu')}  
	\end{aligned}
\end{equation}
To get to the second line from the first we used \ref{LT2}.
If we take
\begin{equation}\label{dualitymap}
	\mu=b'^2 \alpha_+^2\(\frac{\mu'\gamma(b'^2)}{\pi }\)^{1/2}, ~~ N^{0}_{j_a,m_a,\bar{m}_a}=\frac{1}{\pi}, ~~ \mathcal{N}=2\pi
\end{equation}
This precisely matches with the corresponding residues that can be obtained from (\ref{sg3}) if we identify Liouville theory with $(b',\mu')$ with the dual Liouville theory with $(b'',\mu'')$.

\section{Summary and open questions}\label{Summary and open questions}

In this note we discussed a dual description (\ref{winding2}) of the worldsheet CFT of string theory on AdS$_3$ with NS-NS flux (\ref{poincare_1st})\footnote{The pure NS-NS background features a continuum of states that are lifted when an infinitesimal RR flux is turned on, for a  modern discussion of this point see \cite{Cho:2018nfn}.}. In particular, we have found a precise map (\ref{duality_map}) of vertex operators and associated parameters for the winding zero sectors and by explicit calculation of two and three-point functions showed a match. This raises a number of physically interesting questions that we discuss below.

\subsection{Spacetime CFT} \label{ER=EPR}

There are finite energy classical solutions to (\ref{poincare_1st}) that allow the worldsheet to get arbitrarily close to the AdS boundary known as long strings for $k>3$. It has been long speculated that the worldsheet CFT on a long string effectively becomes the spacetime CFT. To make this precise one can quantize the theory (\ref{poincare_1st}) by expanding around such a solution. A careful analysis of the perturbative BRST cohomology in the large negative $\varphi$ region shows that to build the Hilbert space we don't need $\beta,\gamma, b,c$  oscillators \cite{Seiberg:1999xz} and therefore the effective worldsheet CFT that describes the physical states is given by a linear dilaton theory ($\varphi$) coupled with an internal CFT $X$ (for large negative $\varphi$)\footnote{In the language of worldsheet path integration, the path integral over $b,c$ fluctuations cancels the contribution from the path integral over $\beta, \gamma$ fluctuations. However, the identification of worldsheet and spacetime boundary co-ordinates is somewhat subtle.}. 

This suggests that the spacetime CFT is closely related to multiple copies of the linear dilaton theory coupled to the internal CFT $X$ (for large negative $\varphi$), in the same way that AdS$_5$ is dual to the theory on multiple D3 branes. This would result in a spacetime CFT that is the symmetric product orbifold of a linear dilaton theory coupled with the internal CFT $X$. Interestingly, the twisted sectors of the orbifold map to the spectrally flowed states of the long string worldsheet theory. However, this non-interacting picture needs to be corrected away from the weak coupling region $\varphi \rightarrow - \infty$. 

For $k>3$, authors of \cite{Eberhardt:2021vsx} (for a related discussion in the case of $k<3$ see \cite{Martinec:2021vpk, Balthazar:2021xeh})  made a proposal for the complete spacetime CFT - it is the symmetric product orbifold deformed by a marginal operator in twist two-sector. The linear dilaton part  of the marginal operator associated with the deformation takes the following form in the untwisted sector
\begin{equation}
	e^{\sqrt{k-2}\varphi}
\end{equation}
Remarkably this is exactly the slope of the linear dilaton in (\ref{winding2}).

It would be extremely interesting to understand this connection in more detail - obtaining the spacetime CFT directly from the theory on the long strings described in (\ref{winding2}), would shed light on the detailed origin of the AdS/CFT correspondence in these backgrounds.

 \subsection{Stringy phase diagram}

 It would be very interesting to study the full phase diagram of stringy configurations with thermal Euclidean AdS$_3$ boundary conditions using the dual winding description. In higher dimensions, the stable large AdS black hole solution is continuously connected to the unstable small black hole at a minimum macroscopic Schwarzschild radius. It is important to open the question of whether thermal AdS connects smoothly to the small black hole close to the Hagedorn temperature.
In contrast, in three dimensions, the classical gravity BTZ saddle extends to an arbitrarily small Schwarzschild radius, and it is unknown whether it connects to an analog of the unstable small black hole phase when the spatial circle is of order the string scale. Note that the modular S-transform of the boundary dual conformal field theory exchanges the Hagedorn temperature of thermal AdS with that of the string scale horizon BTZ. If there is an analog of the small black hole solution, it must be in a stringy regime, and so may be more accessible using the winding condensate description, as follows.

  In this note the potential in the action (\ref{winding2}) has a non-trivial charge for the winding around the  $\theta$ direction. If we compactify $\xi$ then it would describe TAdS$_3$. We could have equally considered a model in which the deformation potential had a winding charge around the $\xi$ direction instead, which should be equivalent to the BTZ black hole after compactifying $\theta$ direction (as dictated by the modular invariance of the spacetime CFT). 

It is well known that overheated TAdS$_3$, i.e., temperature above the Hagedorn transition $T>T_{H}$, and over cooled BTZ, i.e., temperature $T<4\pi^2/T_{H}$, suffer from instability in winding modes around the non-contractible circle \cite{Berkooz:2007fe, Rangamani:2007fz}. For intermediate temperatures $4\pi^2/T_{H}<T<T_{H}$\footnote{This interval of temperatures exists only for $k>3$.}, we can try to analyze this phenomenon by considering a theory in which both the winding potentials described above are added to the free theory on the cylinder with a relative strength $g$ and follow the worldsheet RG flow to find the effective worldsheet CFT as a function of $T,g$. In particular, it would be very interesting to investigate the existence of a worldsheet CFT that is thermodynamically unstable from the perspective of the target spacetime and potentially connects to BTZ at temperature $4\pi^2/T_H$ and thermal AdS at temperature $T_H$\footnote{More generally we can consider a winding potential corresponding to a generic filling of cycles of the boundary torus of AdS$_3$. We thank S. Minwalla for emphasizing this point.}.
 
 \subsection{Entropy from the worldsheet} 

One of the well-known hard open problems in string theory is the calculation of the sphere partition function from the worldsheet due to unfixed gauge degrees of freedom \cite{Liu:1987nz}. The guess is that the path integral over the zero modes associated with the non-compact directions should play a central role along with these unfixed gauge directions to produce a meaningful result. The key advantage of the dual representation (\ref{winding2}) of the worldsheet CFT for AdS$_3$  is that integration over the radial zero mode effectively inserts a set of winding operators and we can now gauge fix the location of these operators (just like the calculation of three and higher point amplitudes in the usual flat space string theory) to calculate the partition function. Potentially this could provide an explanation of the leading order entropy of the BTZ black hole directly in terms of counting stringy microstates, given that pairs of winding operators correspond to the thermal trace over the Hilbert space in angular quantization\footnote{A different route to computation of thermal entropy would be to use the replica trick of  \cite{Lewkowycz:2013nqa, Faulkner:2013ana} or the technique of orbifolds presented in \cite{Dabholkar:2001if, Witten:2018xfj, Dabholkar:2022mxo}. To this end, it would be interesting to consider the theory obtained by deforming the free theory on the cylinder by winding $\pm \nu$ marginal operators which are uncharged under $J^3, \bar{J}^3$ and understand its relation to $ \mathbb{Z}_\nu$ orbifold of AdS$_3$.}.

  \subsection{A supersymmetric version}
 
 The duality presented in this note is an example of the kind of duality that relates a geometric description in which stringy effects are small, with another description that looks string scale. This is possible because strings are extended objects and geometric features smaller than the string scale are not uniquely probed by them. Mirror symmetry is another example of this type and in fact, might be related to a supersymmetric version of the duality presented here as we explain next.

  If we gauge the symmetry generated by $J^3$ on both sides of the duality presented in this note we obtain the FZZ duality with two-dimensional target space. 
  A supersymmetric version of the duality can be understood in the spirit of Mirror symmetry - each side of the duality is obtained as the IR limit of a supersymmetric field theory and these UV supersymmetric field theories are mirror pair of each other \cite{Hori:2001ax}. This suggests we can use similar techniques to prove the supersymmetric version of the duality presented in this note by clever use of worldsheet RG. That would be the analog of the story presented in \cite{Chen:2021dsw} for AdS$_3$.

\acknowledgments

We would like to thank S. Minwalla and X. Yin for valuable discussions and to thank N. Agia and A. Dei for useful comments. This work was supported in part by DOE grants DE-SC0007870 and DE-SC0021013.

\appendix
\section{Conventions}

We are considering Euclidean worldsheet
\begin{equation}
	\begin{aligned}
		z=\sigma^0+i \sigma^1, ~~ \bar{z}=\sigma^0-i \sigma^1, ~~ \partial_z=\frac{1}{2}(\partial_0-i\partial_1), ~~ \partial_{\bar{z}}=\frac{1}{2}(\partial_0+i\partial_1)
	\end{aligned}
\end{equation}
\begin{equation}
	\begin{aligned}
		g_{z\bar{z}}=g_{\bar{z}z}=\frac{1}{2},~~ g_{zz}=g_{\bar{z}\bar{z}}=0, ~~ \epsilon_{z\bar{z}}=-\epsilon_{\bar{z}z}=\frac{i}{2},~~ \epsilon_{zz}=\epsilon_{\bar{z}\bar{z}}=0
	\end{aligned}
\end{equation}
\begin{equation}
	\begin{aligned}
	d^2 z= d\sigma^0 d\sigma^1, ~~ 	\delta(z,\bar{z})=\delta(\sigma^0)\delta(\sigma^1), ~~ \int d^2 z \ \delta(z,\bar{z})=1
	\end{aligned}
\end{equation}
Stoke's theorem takes the form
\begin{equation}
	\begin{aligned}
	\int d^2 \sigma \ \partial_\mu J^\mu=-i \int dz (J_z dz-J_{\bar{z}}d\bar{z}), ~~~ \mu =0,1
		\end{aligned}
\end{equation}

We'll use the following convention for conserved quantities like stress tensor 
\begin{equation}
	T=-2\pi T_{zz}, ~~ \bar{T}=-2\pi T_{\bar{z}\bar{z}}
\end{equation}

\section{Integral Identities}

We will present the duality relation between certain multi-dimensional integrals. Before that we introduce proper measures by the following formula
\begin{equation}
	\mathcal{D}_n(t)= \prod_{1\leq i<j\leq n} |t_i-t_j|^2, ~~ d^{2n}t=\frac{1}{\pi^nn!} \(\prod_{i=1}^{n} d^2t_i\)
\end{equation}
The first identity that we will use heavily is given by (useful for $n\geq m$) \cite{Baseilhac:1998eq, Fateev:2007qn}
\begin{equation}\label{identity}
	\begin{aligned}
		& ~~~~\int  d^{2n} y \ \mathcal{D}_n(y)  \( \prod_{i=1}^{n} \prod_{j=1}^{n+1+m}|y_i-t_j|^{2p_j}\)\\
		&= \frac{\prod_{j=1}^{n+m+1}\gamma(1+p_j)}{\gamma(1+n+\sum_{j=1}^{n+m+1} p_j)}\prod_{1\leq j<j'\leq n+m+1}|t_j-t_{j'}|^{2p_j+2p_{j'}+2} \\
		&~~~~~~~~~~~~~~~~~~~~~~~~~~~~~~~~
		\int  d^{2m} y \ \mathcal{D}_m(y) \( \prod_{i=1}^{m} \prod_{j=1}^{n+1+m}|y_i-t_j|^{-2p_j-2}\)
	\end{aligned}
\end{equation}
Where
\begin{equation}
	\gamma(x)=\frac{\Gamma(x)}{\Gamma(1-x)}
\end{equation}
The following properties will be important for us
\begin{equation}\label{property}
	\gamma(x)\gamma(1-x)=1, ~~ \gamma(x)\gamma(-x)=-\frac{1}{x^2}
\end{equation}
The second identity takes the following form (useful for $n\geq m$) \cite{Fateev:2007qn}
\begin{equation}\label{identity2}
	\begin{aligned}
		& ~~~~\int  d^{2n} y \ \mathcal{D}_n(y)  \( \prod_{i=1}^{n} \prod_{j=1}^{n+2+m}|y_i-t_j|^{2p_j}\)\\
		&=\prod_{j=1}^{n+m+2}\gamma(1+p_j)\prod_{1\leq j<j'\leq n+m+2}|t_j-t_{j'}|^{2p_j+2p_{j'}+2} \\
		&~~~~~~~~~~~~~~~~~~~~~~~~~~~~~~~~
		\int  d^{2m} y \ \mathcal{D}_m(y) \( \prod_{i=1}^{m} \prod_{j=1}^{n+2+m}|y_i-t_j|^{-2p_j-2}\)
	\end{aligned}
\end{equation}
where it is assumed that $1+n+\sum_{j=1}^{n+m+2} p_j=0$. This is a generalization of the star-triangle relation.

While these two identities are very useful often the measure for the repulsion does not take the form of $\mathcal{D}_n(t)$ and we can use the following identity to obtain recursion relations\cite{Fateev:2007qn}
\begin{equation}\label{recursion}
	\mathcal{D}_n(t)^{-2b^2}=\mathcal{D}_n(t)\frac{\gamma(-n b^2)}{\gamma(-b^2)}\int  d^{2(n-1)} y \ \mathcal{D}_{n-1}(y)  \( \prod_{i=1}^{n-1} \prod_{j=1}^{n}|y_i-t_j|^{-2-2b^2}\)
\end{equation}

\section{Liouville Theory}\label{LT}

To some extent, Liouville theory on disc $D$ of radius $R \to \infty$ is defined by the following Euclidean action (for a review see \cite{Seiberg:1990eb, Teschner:2001rv, Harlow:2011ny})
\begin{equation}\label{LiouvilleLag}
	\begin{aligned}
		& S=\int_{D} d^2 \sigma \[ \frac{1}{\pi}\partial \phi \bar{\partial}\phi +\mu e^{2b \phi}\]+\frac{Q}{\pi}\int_{\partial D}\phi d\theta+2Q^2 \log R, ~~  ~ Q=b+\frac{1}{b}, ~~ b \in (0,1) \\
	\end{aligned}
\end{equation}
with the boundary condition
\begin{equation}
	\phi=-2Q \log |z|+\mathcal{O}(1) ~~~~~~\text{as $|z|\to \infty$}
\end{equation}
The theory is invariant under the following conformal transformation 
\begin{equation}
	\begin{aligned}
		z'=w(z), ~~ \phi(z',\bar{z}')=\phi(z,\bar{z})-\frac{Q}{2}\log |\frac{\partial w(z)}{\partial z}|^2
	\end{aligned}
\end{equation}
generated by the stress tensor
\begin{equation}
\begin{aligned}
	& T=-(\partial \phi)^2+Q\partial^2 \phi\\
\end{aligned}
\end{equation}
associated with the holomorphic conformal anomaly given by $ c=1+6 Q^2$.
The Virasoro primaries are given by\footnote{Here $\alpha$ is in general a complex number.}
\begin{equation}
	V_{\alpha}=e^{2\alpha \phi}, ~~ h_\alpha=\alpha(Q-\alpha)
\end{equation}
$h_\alpha$ is the holomorphic dimension of the operator.
For generic $\alpha$ these are non-degenerate fields, for special values 
\begin{equation}
	\alpha=-\frac{mb}{2}-\frac{n}{2b}, ~~~ m,n=0,1,2,..
\end{equation}
they correspond to degenerate primaries whose Verma module is a small representation of the Virasoro algebra. 

One can study the theory given in (\ref{LiouvilleLag}) in the semi-classical limit defined by $b\to 0$ keeping $\mu b^2$ fixed. Strictly speaking, the action in (\ref{LiouvilleLag}) captures the exact conformal field theory in this limit precisely.  If the operator $V_\alpha$ satisfies  the Seiberg bound
\begin{equation}
	\Re( \alpha )<\frac{1}{2b}
\end{equation}
then in the semi-classical limit near the insertion of such an operator the semi-classical saddle point value of the Liouville field behaves as a free field and can be studied easily. 

Understanding the exact conformal theory beyond the semi-classical limit is more challenging (away from this limit the action in (\ref{LiouvilleLag}) requires corrections as required by the Liouville self-duality that we will describe later). One can still obtain certain exact constraints by performing path integration based on the action in (\ref{LiouvilleLag}). Correlation functions of non-degenerate operators are meromorphic functions and are constrained by the residue at the poles 
\begin{equation}\label{LT1}
	\begin{aligned}
		 &\res_{\sum_a \alpha_a \to (Q-nb)  }\langle V_{\alpha_1}(z_1,\bar{z}_1)V_{\alpha_2}(z_2,\bar{z}_2).. V_{\alpha_m}(z_m,\bar{z}_m) \rangle\\
		 &=\frac{(-\mu)^n}{n!}\prod_{i<j}\[ |z_i-z_j|^{-4 \alpha_i \alpha_j}\] \int \prod_{j=1}^{n}[d^2t_j] \prod_{i<j}\[ |t_i-t_j|^{-4b^2}\]\int \prod_{j=1}^{n}\prod_{k=1}^{m}\[|t_j-z_k|^{-4b\alpha_k}\]
	\end{aligned}
\end{equation}
Here $n$ is a positive integer. Similar equations hold for correlation functions involving degenerate operators (they satisfy the Belavin-Polyakov-Zamolodchikov equation), for example\footnote{This particular formula is taken from  \cite{Fateev:2007qn}.}
\begin{equation}\label{LT2}
	\begin{aligned}
		 &\res_{\sum_a \alpha_a \to (Q-nb+\frac{mb}{2})  }\langle V_{-\frac{mb}{2}}(z,\bar{z})V_{\alpha_1}(0,0)V_{\alpha_2}(1,1) V_{\alpha_3}(\infty,\infty) \rangle\\
		 &=\frac{(-\mu)^n}{n!}|z|^{2mb\alpha_1}|1-z|^{2mb\alpha_2} \int \prod_{j=1}^{n}[d^2t_j] \prod_{i<j}\[ |t_i-t_j|^{-4b^2}\]\int \prod_{j=1}^{n}\[|t_j|^{-4b\alpha_1}|t_j-1|^{-4b\alpha_2} |t_j-z|^{2mb^2} \]
	\end{aligned}
\end{equation}
Where we have omitted a suitable power of the overall scale.

The spectrum of the complete set of Virasoro primary operators of the theory is more subtle. Hints come from studying the associated non-relativistic quantum mechanics obtained from (\ref{LiouvilleLag}) by restricting to field configurations of the form $\phi(\sigma^0)$ in semi-classical limit. A basis of the Hilbert space is given by the eigenstates of the Hamiltonian subjected to regularity near the Liouville wall at $\phi \to +\infty$ - the basis is labeled by $\alpha=Q/2+iP$ for $P>0$. Near $\phi \to -\infty$, the scattering wave functions take the form  $e^{-Q\phi}(V_{\alpha}+R_{L,c}(\alpha) V_{Q-\alpha})$ dictated by the reflection co-efficient $R_{L,c}(\alpha)$ in semi-calssical limit.  

An exact conformal field theory can be constructed in which the delta-function normalizable\footnote{The two-point function takes the form
\begin{equation}
    \langle O_{\alpha}(z,\bar{z}) O_{\alpha'}(0,0)\rangle= \frac{2\pi \delta(\alpha-\alpha')}{z^{2h_\alpha} \bar{z}^{2h_\alpha}}
\end{equation}}
primaries (i.e, states) take the form $$O_{\alpha}= \frac{1}{2}R^{-1/2}_L(\alpha) (V_{\alpha}+R_L(\alpha) V_{Q-\alpha}), ~~ \alpha=Q/2+iP, ~~ P>0$$where\footnote{We identify operators according to \begin{equation}
	V_{Q-\alpha}=R_L(Q-\alpha) V_\alpha
\end{equation} }
\begin{equation}
	R^{-1}_L(\alpha)=\frac{b^2 (\pi \mu \gamma(b^2))^{(2\alpha-Q)/b}}{\gamma(2\alpha/b-1-1/b^2)\gamma(2\alpha b-b^2)}
\end{equation}
The operator product expansion of these primaries converges. In the conformal field theory, degenerate operators represent null states. The exact three-point function is given by the DOZZ formula \cite{Dorn:1994xn, Zamolodchikov:1995aa}
\begin{equation}
	\langle V_{\alpha_1}(0,0)V_{\alpha_2}(1,1) V_{\alpha_3}(\infty, \infty) \rangle = \left[\pi \mu \gamma(b^2) b^{2-2b^2}\right]^{\frac{Q-\sum_{k=1}^{3}\alpha_k}{b}} \frac{ \Upsilon_{b}'(0) \prod_{k=1}^{3} \Upsilon_{b}(2\alpha_k) }{\Upsilon_{b}(\sum_{k=1}^{3} \alpha_k - Q) \prod_{k=1}^3 \Upsilon_{b}(\sum_{k=1}^{3} \alpha_k -2\alpha_k)}
\end{equation}
Here $\Upsilon_b(x)$ is given by 
\begin{equation}
\label{logupsilon}
\log\Upsilon_b(x)=\int_0^\infty\frac{dt}{t}\left[(Q/2-x)^2 e^{-t}-\frac{\sinh^2((Q/2-x)\frac{t}{2})}{\sinh{\frac{tb}{2}}\sinh{\frac{t}{2b}}}\right]\qquad 0<\mathrm{Re}(x)< Q.
\end{equation}
The integral representation is valid for $ 0<\mathrm{Re}(x)< Q$. $\Upsilon_b(x)$
has an analytic continuation to an entire function of $x$.  The three-point function (operator insertions are unchanged under the duality) in the theory is invariant under the Liouville self-duality duality
\begin{equation}\label{LTD}
	b'=\frac{1}{b}, ~~ \pi \mu'\gamma(b'^2)=(\pi \mu\gamma(b^2))^{1/b^2}
\end{equation}

\bigskip

\providecommand{\href}[2]{#2}\begingroup\raggedright\endgroup

\end{document}